\documentclass[sigconf,natbib=true,anonymous=false]{acmart}
\setcopyright{none}
\renewcommand\footnotetextcopyrightpermission[1]{}

\usepackage[hang,flushmargin]{footmisc}

\usepackage{xspace}
\usepackage{graphicx}
\usepackage{booktabs} 
\usepackage{xspace}
\usepackage{multirow}
\usepackage{caption}
\usepackage{subcaption}

\newcommand{\ignore}[1]{}
\newcommand{\cls}{\texttt{[CLS]}\xspace}

\newcommand{\monoBERT}{monoBERT\xspace} 
\newcommand{\monoELECTRA}{monoELECTRA\xspace} 
\newcommand{\minilm}{MiniLM\xspace} 
\newcommand{\electrabase}{ELECTRA\textsubscript{base}\xspace} 
\newcommand{\electralarge}{ELECTRA\textsubscript{large}\xspace}

\newcommand{\msmarco}{MS MARCO\xspace} 
\newcommand{\beir}{BEIR\xspace}

\newcommand{\dtok}{$D_{\text{tok}}$\xspace}

\newcommand{\cross}[0]{\textcolor{black}{\textbf{$\times$}}}
\newcommand{\tick}[0]{\textcolor{black}{\textbf{\checkmark}}}
\usepackage{amsmath}

\AtBeginDocument{%
  \providecommand\BibTeX{{%
    \normalfont B\kern-0.5em{\scshape i\kern-0.25em b}\kern-0.8em\TeX}}}

\setcopyright{acmcopyright}
\copyrightyear{2018}
\acmYear{2018}
\acmDOI{XXXXXXX.XXXXXXX}

\begin{document}

\title{Improving Out-of-Distribution Generalization of Neural Rerankers with Contextualized Late Interaction} 

\author{
Xinyu Zhang$^{*}$, 
Minghan Li$^{*}$,
Jimmy Lin
}
\thanks{$^*$ Equal Contribution}

\affiliation{\vspace{0.1cm}
David R. Cheriton School of Computer Science, University of Waterloo \country{Canada}
}

\begin{abstract}
Recent progress in information retrieval finds that embedding query and document representation into multi-vector yields a robust bi-encoder retriever on out-of-distribution datasets.
In this paper, we explore whether late interaction, the simplest form of multi-vector, is also helpful to neural \textit{rerankers} that only use the \texttt{[CLS]} vector to compute the similarity score. 
Although intuitively, the attention mechanism of rerankers at the previous layers already gathers the token-level information,
we find adding late interaction still brings an extra 5\% improvement in average on out-of-distribution datasets, with little increase in latency and no degradation in in-domain effectiveness.
Through extensive experiments and analysis,
we show that the finding is consistent across different model sizes and first-stage retrievers of diverse natures,
and that the improvement is more prominent on longer queries.
\end{abstract}

\maketitle

\section{Introduction}

The two-stage retrieve-then-rerank pipeline has been the \textit{de facto} design for many information retrieval systems.
With the advancement in pre-trained language models, these retrieval systems also benefit from the rich semantics in the contextualized representations which could be fine-tuned for measuring the similarity between queries and documents.
Commonly, the $\texttt{[CLS]}$ token vector at the last layer is often chosen to be the sequence-level representation.
However, neural retrievers that only use the \texttt{[CLS]} vector might be less robust on out-of-distribution (OOD) datasets as some of the token-level granularity might not be captured.
Therefore, methods such as further pre-training or adding token-level interaction have been applied to improve the OOD generalization of the neural retrievers.
Among them, 
late interaction models~\cite{colbert,coil}, also known as the multi-vector retrievers, 
strikes a perfect balance between the in-domain and out-of-domain effectiveness among neural retrievers.
This is usually credited to its design which takes the last layer of contextualized token embeddings to compute the final similarity other than just using the \texttt{[CLS]} vector. Given its powerful design, in this paper, we raise the following question:

\textit{Can neural rerankers that only use the \texttt{[CLS]} vector for computing similarity scores also benefit from adding the late interaction?}

Intuitively, many would argue that the attention mechanism at the previous layers already gathers the token-level interaction between the query and documents.
However, in this paper, we show that late interaction at the last layer actually brings ``free'' OOD capacity to rerankers. 
As shown \autoref{tab:overview}, after adding late the interaction, the averaged nDCG@10 on BEIR is improved by 5\% (from 0.467 to 0.491),
while the in-domain score (MRR@10 on MS MARCO) is not affected and the search latency is only slightly increased.
We also show that this improvement is orthogonal to the better OOD capacity brought by larger model size,
and consistent when reranking candidates from all categories of retrievers.

\begin{table}[]
    \centering
    \resizebox{0.9\columnwidth}{!}{
    \begin{tabular}{c c c c}
    \toprule
    \begin{tabular}[r]{@{}c@{}} \textbf{Add Late}\\\textbf{Interaction?}\end{tabular} 
    & \begin{tabular}[c]{@{}c@{}} \textbf{MS MARCO}\\{MRR@10}\end{tabular} & \begin{tabular}[c]{@{}c@{}} \textbf{BEIR Avg}\\{nDCG@10}\end{tabular} & \begin{tabular}[c]{@{}c@{}} \textbf{Search}\\\textbf{Latency}\end{tabular}\\
    \midrule
        \cross & 0.390 & 0.467 & 1.18s \\
        \tick & 0.392 & 0.491 & 1.28s\\
    \bottomrule
    \end{tabular}
    }
    \caption{
    The in-domain score (MRR@10 on MS MARCO), OOD score (averaged nDCG@10 on BEIR), and search latency of rerankers w/o and w/ adding late interaction. 
    Rerankers are initialized from \minilm. 
    }
    \vspace{-2em}
    \label{tab:overview}
\end{table}

\section{Related Work}
\citet{monobert} was one of the first work on cross-encoders, which serve to rerank a subset of documents returned from the ``first-stage'' retrievers.
It considers retrieval as a classification task, 
and uses Transformer encoders following the formulation of the next sentence prediction (NSP) pretraining task in BERT, 
where only the \cls vector is used in classifying the (query, document) pair and computing the relevant score.
Afterwards,
CEDR~\cite{cedr} also proposes to integrate token information rather than use only \cls,
but it processes token representations at all Transformer layers using the pre--BERT neural rerankers~\cite{knrm, drmm, pacrr},
which is more complex in structure and adds higher computational overhead.

Recent works on first-stage retrieval have demonstrated the effectiveness of adding sparse information into dense retrieval. 
\citet{spar}
The combination of the token information and dense \cls vector could also be done explicitly,
by either adding the scores computed from \cls and token information, or concatenated aggregated token vectors to the \cls vector~\cite{coil, aggretriever}.
The multi-vector dense models could also be viewed under this category,
where the token representation vectors jointly contribute to the relevancy computation along with the \cls vector~\cite{colbert,li2022citadel}.

\begin{table*}[t]
\resizebox{\textwidth}{!}{
\begin{tabular}{lc|c|c|ccccccccccccc}
\toprule
\multirow{3}{*}{\textbf{Backbone}} & \multirow{3}{*}{\textbf{\begin{tabular}[r]{@{}c@{}}Add Late\\ Interaction?\end{tabular}}} & \multirow{3}{*}{\begin{tabular}[c]{@{}c@{}} \textbf{MS}\\\textbf{MARCO} \\ (MRR@10) \end{tabular}} & \multicolumn{14}{c}{\textbf{BEIR} (nDCG@10)} \\
 &  &  & \textbf{Avg} & \begin{tabular}[c]{@{}c@{}}TREC-\\ COVID\end{tabular}  &  \begin{tabular}[c]{@{}c@{}}NF\\ Corpus \end{tabular} & NQ & \begin{tabular}[c]{@{}c@{}}Hotpot\\ QA\end{tabular} & FiQA & \begin{tabular}[c]{@{}c@{}}Argu\\Ana\end{tabular} & \begin{tabular}[c]{@{}c@{}}Touche-\\ 2020\end{tabular} & Quora & \begin{tabular}[c]{@{}c@{}}DB\\Pedia\end{tabular}  & \begin{tabular}[c]{@{}c@{}}SCI\\DOCS\end{tabular}  & FEVER & \begin{tabular}[c]{@{}c@{}}Climate-\\ FEVER\end{tabular} & \begin{tabular}[c]{@{}c@{}}Sci\\Fact\end{tabular} \\
 \midrule
\multirow{2}{*}{\textbf{MiniLM}} & \cross & 0.390 & 0.467 & 0.699 & 0.355 & 0.504 & 0.620 & 0.359 & 0.335 & 0.308 & 0.722 & 0.426 & 0.151 & 0.754 & 0.164 & 0.679 \\
& \tick & 0.392 & 0.491 & 0.705 & 0.349 & 0.501 & 0.673 & 0.360 & 0.527 & 0.324 & 0.784 & 0.424 & 0.155 & 0.723 & 0.172 & 0.691 \\
\midrule
\multirow{2}{*}{\textbf{\electrabase}} & \cross &  0.400 & 0.481 & 0.727 & 0.362 & 0.523 & 0.660 & 0.389 & 0.291 & 0.317 & 0.773 & 0.436 & 0.152 & 0.748 & 0.112 & 0.669 \\
 & \tick & 0.402 & 0.494 & 0.736 & 0.368 & 0.527 & 0.714 & 0.401 & 0.443 & 0.320 & 0.690 & 0.449 & 0.162 & 0.740 & 0.152 & 0.715 \\
 \midrule
\multirow{2}{*}{\textbf{\electralarge}} & \cross &  0.413 & 0.507 & 0.801 & 0.380 & 0.559 & 0.733 & 0.453 & 0.250 & 0.339 & 0.772 & 0.468 & 0.181 & 0.791 & 0.149 & 0.719 \\
 & \tick & 0.413 & 0.524 & 0.786 & 0.378 & 0.559 & 0.735 & 0.457 & 0.436 & 0.335 & 0.800 & 0.460 & 0.182 & 0.769 & 0.179 & 0.733 \\
 \bottomrule
\end{tabular}
}
\caption{
MRR@10 on \msmarco and nDCG@10 scores on \beir.
Rerankers are intialized from \minilm, \electrabase, and \electralarge.
Results on \beir rerank the top-1k passages from BM25.
}
\label{tab:modelsize}
\vspace{-2em}
\end{table*}

\section{Methods}
In this section, we introduce \monoBERT, the reranker we used in this work, and how we apply late interaction on the reranker.

\subsection{\monoBERT}
\label{sec:methodmonobert}
\monoBERT~\cite{monobert} is one of the first works that applied pretrained transformers in passage retrieval. 
The model is fed with concatenated query $q$ and document $d$ and computes relevance scores $s_{q, d}$ from the \cls representation on the last layer of the Transformer encoder.
We borrow the following formulations from \citet{book} and \citet{squeezewater}:
\begin{align}
\label{eq:monobert}
    s_{m}(q, d) = s_{\monoBERT}(q, d) = T_{\cls}W + b
\end{align}
where $T_\cls \in \mathbf{R}^{D}$ is the \cls representation on the final layer,
and $W \in \mathbf{R}^{D \times 1}$ and $b \in \mathbf{R}$ are the weight and bias for classification. 

\smallskip
It is worth noting that we use the term \monoBERT as a reference to the \textit{model architecture}, rather than the model being initialized from BERT. 
Instead, the model may use other pretrained Transformers checkpoint, or \textit{backbone} following the standard parlance in the community.
For example, \citet{squeezewater} initialized it from ELECTRA~\cite{electra} and referred to the model as \monoELECTRA.
We do not follow this naming schema as multiple backbones in different scales are used. 
To avoid confusion,
we always refer to the model as \monoBERT, and specify the backbone that it is initialized from.

We train all models using the LCE loss~\cite{LCE}, which has been proven to be more effective~\cite{LCE, squeezewater, simlm}.
Detailed configuration will be introduced in Section~\ref{sec:setup}.

\subsection{Late Interaction}
In this work, we use the simplest version of late interaction proposed by \citet{colbert}.
While there are more variants on multi-vector retrievers, they are designed mostly to mitigate the huge index storage footprint, which is not a concern in the reranking stage. 

We first obtain the representation of each token in the query $q$ and document $d$: 
\begin{align}
\vspace{-1em}
\label{eq:rep_lexical}
    v_{q_i} = T_{q_i}W_{q_i} + b_{q_i}; \quad 
    v_{d_j} = T_{d_j}W_{d_j} + b_{d_j};
\end{align}
where $q_i$ and $d_j$ represent the $i$-th token of query $q$ and the $j$-th token of document $d$, respectively. 
Similar to \autoref{eq:monobert},
$T \in \mathbf{R}^{D}$ is the representation of each token on the final layer. 
$W \in \mathbf{R}^{D \times D_{tok}}$ and $b \in \mathbf{R}^{D_{tok}}$ are the weight and bias in a projection layer,
which projects the $T_{tok}$ to a lower dimension $D_{tok} < D$.

Eq.~\ref{eq:rep_lexical} is the defualt setting in all experiments. In Section~\ref{sec:dimension}, we investigate another variant where the token representation are not projected:
\begin{align}
\vspace{-2em}
\label{eq:rep_lexical_noproj}
    v_{q_i} = T_{q_i}; \quad  v_{d_j} = T_{d_j}
\end{align}

With token representaions $v_{q_i}$ and $v_{d_j}$,
the late interaction then computes the token interaction scores score by summing up the inner product between all tokens in queries and documents:
\begin{align}
\label{eq:maxsim}
\vspace{-1em}
    s_{l}(q, d) = s_{late-interaction}(q, d) = \sum_{q_i} \max_{d_j} (v_{q_i}^T v_{d_j})
\end{align}

It shares the same formulation with the first-stage retrievers,
and only differ in that the token representation $T_{q_i}$ and $T_{d_j}$ are computed joinly with both query and document information,
whereas in retrievers, 
they are computed independently with each other,
with $T_{q_i}$ perceiving no information from document $d$ and vice versa. 

At training time, we train $s_{m}$ and $s_{l}$ (or $s'_l$) independently:
\begin{align}
\vspace{-1em}
l &= L_1(s_{m}(q, d_{0}), ..., s_{m}(q, d_{n})) + L_2(s_{l}(q, d_{0}), ..., s_{l}(q, d_{n}))
\end{align}
where the loss functions $L_1$ and $L_2$ could be same or different,
where both use LCE~\cite{LCE,squeezewater} in this work. 
At inference time, we sum the two scores as the final relevance score, i.e., $s_{final} = s_{m} + s_{l}$.\footnote{
We explored adding weighting terms for $s_{m}$ and $s_{c}$, but only observed marginal gains.
Thus we report the simplest formulation here.
}

\begin{table*}[t]
\resizebox{\textwidth}{!}{
\begin{tabular}{lc|c|ccccccccccccc}
\toprule
\multirow{4}{*}{\textbf{First Stage}} & \multirow{4}{*}{\textbf{\begin{tabular}[r]{@{}c@{}}Add Late\\ Interaction?\end{tabular}}} & \multicolumn{14}{c}{\textbf{BEIR} (nDCG@10)} \\
 &  & \textbf{Avg} & \begin{tabular}[c]{@{}c@{}}TREC-\\ COVID\end{tabular}  &  \begin{tabular}[c]{@{}c@{}}NF\\ Corpus \end{tabular} & NQ & \begin{tabular}[c]{@{}c@{}}Hotpot\\ QA\end{tabular} & FiQA & \begin{tabular}[c]{@{}c@{}}Argu\\Ana\end{tabular} & \begin{tabular}[c]{@{}c@{}}Touche-\\ 2020\end{tabular} & Quora & \begin{tabular}[c]{@{}c@{}}DB\\Pedia\end{tabular}  & \begin{tabular}[c]{@{}c@{}}SCI\\DOCS\end{tabular}  & FEVER & \begin{tabular}[c]{@{}c@{}}Climate-\\ FEVER\end{tabular} & \begin{tabular}[c]{@{}c@{}}Sci\\Fact\end{tabular} \\

 \midrule
\multicolumn{16}{c}{\textbf{\textit{Sparse}}} \\ 
\multirow{2}{*}{\textbf{BM25}} & \cross &   0.467 & 0.699 & 0.355 & 0.504 & 0.620 & 0.359 & 0.335 & 0.308 & 0.722 & 0.426 & 0.151 & 0.754 & 0.164 & 0.679 \\
 & \tick &  0.491 & 0.705 & 0.349 & 0.501 & 0.673 & 0.360 & 0.527 & 0.324 & 0.784 & 0.424 & 0.155 & 0.723 & 0.172 & 0.691 \\
 \midrule
\multirow{2}{*}{\textbf{uniCOIL}} & \cross &   0.426 & 0.711 & 0.337 & 0.556 & 0.576 & 0.271 & 0.335 & 0.277 & 0.727 & 0.426 & 0.152 & 0.375 & 0.116 & 0.680 \\
 & \tick &  0.452 & 0.713 & 0.328 & 0.552 & 0.625 & 0.272 & 0.555 & 0.285 & 0.784 & 0.423 & 0.156 & 0.360 & 0.128 & 0.691 \\
 \midrule
\multirow{2}{*}{\textbf{SPLADE}} & \cross &   0.469 & 0.706 & 0.336 & 0.563 & 0.617 & 0.362 & 0.320 & 0.278 & 0.728 & 0.434 & 0.152 & 0.758 & 0.160 & 0.682 \\
 & \tick &  0.492 & 0.699 & 0.330 & 0.560 & 0.671 & 0.361 & 0.526 & 0.288 & 0.786 & 0.432 & 0.157 & 0.717 & 0.173 & 0.691 \\
 \midrule
\multicolumn{16}{c}{\textbf{\textit{Single-vector Dense}}} \\ 
\multirow{2}{*}{\textbf{DPR (Wiki)}} & \cross &   0.451 & 0.699 & 0.335 & 0.571 & 0.600 & 0.341 & 0.333 & 0.285 & 0.523 & 0.433 & 0.154 & 0.753 & 0.175 & 0.662 \\
 & \tick &  0.472 & 0.715 & 0.330 & 0.568 & 0.643 & 0.339 & 0.524 & 0.296 & 0.557 & 0.432 & 0.156 & 0.721 & 0.180 & 0.673 \\
 \midrule
\multirow{2}{*}{\textbf{DPR (MS)}} & \cross &   0.474 & 0.737 & 0.334 & 0.562 & 0.613 & 0.364 & 0.336 & 0.278 & 0.718 & 0.434 & 0.153 & 0.771 & 0.181 & 0.677 \\ 
 & \tick &  0.495 & 0.738 & 0.329 & 0.557 & 0.655 & 0.364 & 0.528 & 0.287 & 0.782 & 0.434 & 0.156 & 0.738 & 0.186 & 0.687 \\ 
 \midrule
 \multirow{2}{*}{\textbf{ANCE}} & \cross &   0.471 & 0.724 & 0.331 & 0.554 & 0.594 & 0.360 & 0.338 & 0.285 & 0.717 & 0.419 & 0.155 & 0.781 & 0.192 & 0.676 \\
 & \tick &  0.493 & 0.740 & 0.327 & 0.550 & 0.626 & 0.363 & 0.529 & 0.291 & 0.781 & 0.418 & 0.157 & 0.750 & 0.192 & 0.687 \\
 \midrule
\multirow{2}{*}{\textbf{TCT~ColBERT}} & \cross &   0.470 & 0.719 & 0.336 & 0.564 & 0.620 & 0.360 & 0.319 & 0.281 & 0.714 & 0.437 & 0.154 & 0.767 & 0.170 & 0.676 \\ 
 & \tick &  0.494 & 0.725 & 0.330 & 0.560 & 0.665 & 0.360 & 0.524 & 0.291 & 0.780 & 0.438 & 0.157 & 0.733 & 0.177 & 0.689 \\  
 \midrule 
\multirow{2}{*}{\textbf{TAS-B}} & \cross &   0.472 & 0.714 & 0.338 & 0.565 & 0.623 & 0.361 & 0.333 & 0.281 & 0.727 & 0.436 & 0.153 & 0.760 & 0.167 & 0.680 \\
 & \tick &  0.494 & 0.713 & 0.331 & 0.560 & 0.670 & 0.358 & 0.527 & 0.292 & 0.787 & 0.435 & 0.157 & 0.729 & 0.176 & 0.689 \\
 \midrule
\multicolumn{16}{c}{\textbf{\textit{Multi-vector Dense}}} \\ 
\multirow{2}{*}{\textbf{ColBERT v2}} & \cross &   0.467 & 0.707 & 0.333 & 0.564 & 0.621 & 0.360 & 0.316 & 0.278 & 0.716 & 0.434 & 0.152 & 0.756 & 0.156 & 0.679 \\
 & \tick &  0.493 & 0.709 & 0.327 & 0.560 & 0.672 & 0.361 & 0.525 & 0.291 & 0.780 & 0.431 & 0.157 & 0.724 & 0.178 & 0.691 \\
\bottomrule 
\end{tabular}
}
\caption{nDCG@10 scores on \beir, reranking the top-1k passages from each first-stage retriever.}
\vspace{-2em}
\label{tab:retrievers}
\end{table*}

\section{Experimental Setup}
\label{sec:setup} 
For in-domain evaluation, we train and evaluate the cross-encoders on \msmarco~\cite{msmarco},
where the queries are prepared from Bing search log, and the collection contains paragraph from general Web.
It contains 8.8M passages, over 500k training pairs and has 6980 queries in the small development set.
For out-fo-domain evaluation, we evaluate on 13 datasets of \beir~\cite{beir} due to license reasons.

We implement the model based on Capreolus~\cite{capreolus1,capreolus2},
an IR toolkit for end-to-end neural ad hoc retrieval that focuses on cross-encoders.
We use the logic of training and inference in Capreolus for \msmarco,
and implement the inference on \beir based on the sample script it provides.\footnote{\url{https://github.com/beir-cellar/beir}}

All training configurations follow~\citet{squeezewater}: 
We train \msmarco on 30k steps with learning rate $1e-5$ and batch size~16.
We use linear warmup on the first 3k steps, then linearly decay the learning rate on the following steps. 
On all experiments, we train \monoBERT using the LCE loss~\cite{LCE, squeezewater} with the number of negative samples to be 7.

At the inference stage,
we always rerank top-$1k$ results from the first-stage retrievers.
On \msmarco, we use TCT-ColBERT~\cite{tct-colbert} as the retriever following~\citet{squeezewater}.
On \beir, we use a list of retrievers that covers the categories of sparse, single- and multi-vector dense retrievers.
Retrievers results are produced using one of Pyserini~\cite{pyserini}, BEIR~\cite{beir} repository, and ColBERT~\cite{colbert} repository.\footnote{\url{https://github.com/stanford-futuredata/ColBERT}}
We will provide more details along with the code release.

We experimented with three backbones in this work: 
\minilm~\cite{minilm}, \electrabase~\cite{electra}, and \electralarge~\cite{electra}.
All models are available on HuggingFace~\cite{huggingface}.
\footnote{\texttt{microsoft/MiniLM-L12-H384-uncased}, \texttt{google/electra-base-discriminator}, and \texttt{google/electra-large-discriminator} on HuggingFace.}

\section{Results and Analysis}
\label{sec:results}
\autoref{tab:overview} provides a preview on the effect of adding late interaction on top of rerankers,
where we observed a ``free'' gain on OOD capacity by adding the late interaction.
In this section, we examine our findings in multiple settings,
showing its consistency over different model sizes 
and first-stage retrievers of different natures.

\subsection{Model Size}

Previous papers found that the generalization ability could depend on the model scale.
Specifically, models with a more extensive set of parameters can better generate on unseen distribution\cite{gtr}.
This leads to our question:
\textit{does late interaction remain helpful in improving OOD capacity when initialized from backbones in larger scales and with potentially better generalization ability?}

The answer is that the contribution of late information barely depends on the model size.
\autoref{tab:modelsize} shows both in-domain (on \msmarco) and out-of-domain (on \beir) scores on rerankers without and with adding late interaction after the final layer,
where the rerankers are initialized from three different sizes of backbone: \minilm, \electrabase and \electralarge, where the size of later two models are roughly $2\times$ and $4\times$ of \minilm. 

\smallskip
While we observe higher average scores on \beir as the model size increases,
which echos the previous finding that better generalization ability could be gained as the model scales up, 
the relative improvement brought by token information is similar across the backbones.
On both \electrabase and \electralarge,
adding late interaction drastically improves the average nDCG@10 on \beir,
from 0.474 to 0.502 with \electrabase and from 0.507 to 0.524 with \electralarge.
Additionally, the in-domain scores on the other two backbones are not affected as well,
suggesting that the ``free'' gain is consistent over different model sizes.

\begin{table*}[t]
\resizebox{0.95\textwidth}{!}{
\begin{tabular}{l|c|ccccccccccccc}
\toprule
\multirow{4}{*}{\textbf{\begin{tabular}[r]{@{}c@{}}\\\end{tabular}}} & \multicolumn{13}{c}{\textbf{BEIR} (nDCG@10)} \\
 & \textbf{Avg} & \begin{tabular}[c]{@{}c@{}}TREC-\\ COVID\end{tabular}  &  \begin{tabular}[c]{@{}c@{}}NF\\ Corpus \end{tabular} & NQ & \begin{tabular}[c]{@{}c@{}}Hotpot\\ QA\end{tabular} & FiQA & \begin{tabular}[c]{@{}c@{}}Argu\\Ana\end{tabular} & \begin{tabular}[c]{@{}c@{}}Touche-\\ 2020\end{tabular} & Quora & \begin{tabular}[c]{@{}c@{}}DB\\Pedia\end{tabular}  & \begin{tabular}[c]{@{}c@{}}SCI\\DOCS\end{tabular}  & FEVER & \begin{tabular}[c]{@{}c@{}}Climate-\\ FEVER\end{tabular} & \begin{tabular}[c]{@{}c@{}}Sci\\Fact\end{tabular} \\
 \midrule
Full late interaction &  0.491 & 0.705 & 0.349 & 0.501 & 0.673 & 0.360 & 0.527 & 0.324 & 0.784 & 0.424 & 0.155 & 0.723 & 0.172 & 0.691 \\
 \multicolumn{1}{r|}{w/o EM} & 0.489 & 0.705 & 0.350 & 0.502 & 0.667 & 0.357 & 0.517 & 0.323 & 0.792 & 0.425 & 0.153 & 0.718 & 0.166 & 0.680 \\ 
\bottomrule 
\end{tabular}
}
\caption{
nDCG@10 scores on \beir, where the rerankers are initialized from \minilm and rerank the top-1k passages BM25.
Both results use late interaction, but the second row removes the interaction between the exact-match (EM) tokens.
}
\vspace{-2em}
\label{tab:nonem}
\end{table*}

\begin{table}[t]
\resizebox{0.9\columnwidth}{!}{
\begin{tabular}{ll|c|c}
\toprule
& \multirow{2}{*}{\begin{tabular}[l]{@{}l@{}}Projected Token Dimension \\(\dtok) \end{tabular}} & \begin{tabular}[c]{@{}c@{}} \textbf{MS} \textbf{MARCO}\end{tabular} & \textbf{\beir} \\
& & MRR@10 & \multicolumn{1}{c}{nDCG@10} \\
\midrule
(1) & \dtok$=1$ & 0.3920 & 0.4890 \\
(2) & \dtok$=32$ & 0.3920 & 0.4914 \\ 
(3) & \dtok$=128$ & 0.3920 & 0.4910 \\ 
(4) & \dtok$=384$ & 0.3900 & 0.4911 \\
\bottomrule
\end{tabular}
}
\caption{
MRR@10 of \msmarco and nDCG@10 of \beir, using different dimensions of token representation (\dtok in Eq.~\ref{eq:rep_lexical}).
We report scores to 4 digits here as the values are close in all conditions.
}
\vspace{-2em}
\label{tab:dimension}
\end{table}

\subsection{First-stage Retriever}

\noindent
In recent years, many first-stage retrievers are proposed, which are mainly categorized into dense and sparse retrievers, and the dense retrievals can be further categorized into single- and multi-vector dense retrievers.
After adding the late interaction,
the model architecture of the reranker is now the most similar to the multi-vector dense retrievers.
We thus explore whether this architecture similarity would indicate lower gains while reranking multi-vector retrievers,
and in a broader sense, whether late interaction gives higher improvement when reranking any specific type of retrievers.

\autoref{tab:retrievers} shows the reranking results on BEIR on an extensive list of retrievers, covering all three categories above. 
Looking at the averaged nDCG@10 on \beir,
we do not observe any clear preference for any specific category of retrievers.
That is, we found that late interaction consistently improves the OOD capacity when using retrievers in different natures,
bringing a similar degree of improvement of 0.02-0.03 on average.

\subsection{Token Dimensions}
\label{sec:dimension}
In first-stage retrievals,
it is common to project the token representation into lower dimensions as restricted by indexing storage space and search efficiency.
However, the representations are computed on the fly for rerankers without any index storage.
That is, in the context of rerankers, using representations in higher dimensions does not bring any additional storage cost and only minor searching latency. 
We thus examine whether using higher token dimensions \dtok is helpful.

Results are shown in \autoref{tab:dimension}, where row~(2) corresponds to the BM25 results reported in \autoref{tab:retrievers}.
Comparing rows~(1--4), we found that 
the token dimensions have little impact on effectiveness on \beir:
on row~(1), using $dim=1$ already obtains $0.4890$ on average \beir, while increasing the dimension to $dim=32$ and onwards only provides marginal improvement.

\section{Analysis}
In this section, we present our analysis results of investigating how late interaction improves the OOD capacity of rerankers from two perspectives.
Specifically,
the OOD improvement brought by late interaction 
(1) is not brought by the exact-matched lexical signals at all;
and (2) is more prominent on longer queries.

\subsection{Exact-matched Lexical Signal}
Given that sparse retrievers are in general more generalizable in OOD scenario~\cite{beir}, 
an intuitive hypothesis to the OOD improvement brought by the late interaction is that it considers the exact-matched tokens between queries and documents.

To test the hypothesis,
we train a reranker where interactions between exact-match terms are removed from the late interaction,
and only preserved the ones between \textit{non-exact-matched} ones.
This is formulated as: 
\begin{align}
\vspace{-1em}
\label{eq:nonem}
    s'_{l}(q, d) = \sum_{q_i} \max_{d_j \neq q_i} (v_{q_i}^T v_{d_j})
\vspace{-1em}
\end{align}
Results are shown in \autoref{tab:nonem}.
The first row is the same as the BM25 results in \autoref{tab:retrievers},
whereas the second row shows the results removing the exact-match token interaction. 
To our surprise, 
the average nDCG@10 on \beir only drops slightly after removing the exact-match terms. 
This clearly shows that OOD capacity improvement brought by the late interactions is not through the embedded exact-match signals but other factors instead.

\begin{figure}[t]
    \centering
    \includegraphics[width=0.49\columnwidth, trim={1em 0 1em 2.3em}, clip]{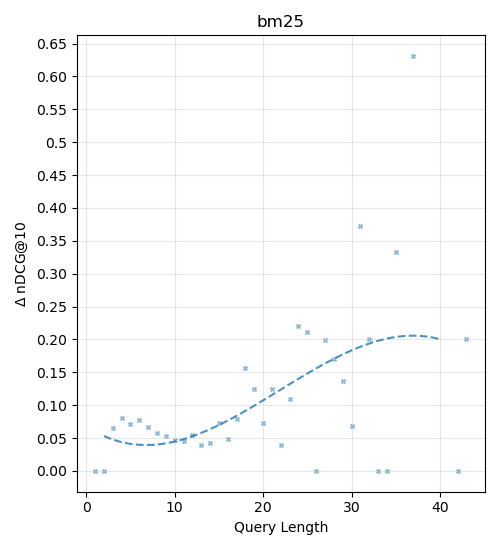}
    \includegraphics[width=0.49\columnwidth, trim={1em 0 1em 2.3em}, clip]{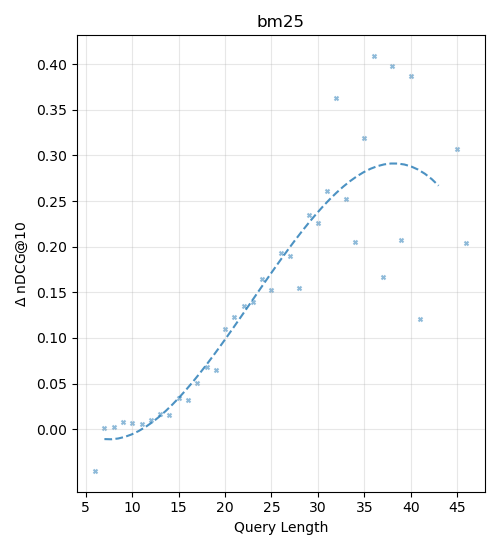}

    \vspace{-1.2em}
    \caption{
        nDCG@10 improvement from late interaction on queries over different lengths. 
        Each point represents the average of nDCG@10 improvements over the query of the corresponding length.
        The line is the least square polynomial fit of the points.
        (Left: Quora, Right: HotpotQA)
    }
    \vspace{-1em}
    \label{fig:qlen}
\end{figure}

\subsection{Query Length}
We explored many aspects of query properties on finding the groups of queries that benefit the most from the late interaction,
including the challenging level of the query,
the completeness of the positive document,
the ranking similarity among different retrievers, and so on. 
In the end, we found that query length is the most prominent indicator of the per-query improvement. 

\autoref{fig:qlen} plots the distribution of nDCG@10 improvement by late interaction according to the query length on Quora and HotpotQA.\footnote{Length determined as the number of query tokens delimited by whitespace.}
Specifically, each point represents the average of nDCG@10 improvements over the query of the same length (same coordinate on x-axis).
We additionally plot an approximated polynomial line based on the points to better reveal the relationship between the query length and nDCG@10 improvement. 

On both datasets, we observe a clear tendency that late interaction brings higher improvement on longer queries.
Here we report results using BM25 as the retriever, while the observation is similar when reranking candidates from other retrievers.

\section{Conclusion}
In this work,
we presented our finding that adding late interaction to existing rerankers brings visible improvement to out-of-distribution capacity without any degradation on in-domain effectiveness,
even though the reranker already processes the token interaction via the attention mechanism at previous layers. 
Extensive experiments on different model sizes and first-stage retrievers show that this improvement is consistent, and according to our analysis, the improvement is more prominent on longer queries.
Our findings suggest that boiling all information into the \texttt{[CLS]} token may not be the optimal choice for neural rerankers,
and more studies are required to better explore its capacity.

\bibliography{acmart}
\bibliographystyle{ACM-Reference-Format}

\end{document}